\documentclass[usegraphicx]{mn2e}

\title[High-performance $N$-body simulations using wavelets]
      {{\boldmath $N$}-body simulations with\\
       two-orders-of-magnitude higher performance\\
       using wavelets}

\author[A. B. Romeo, C. Horellou and J. Bergh]
       {Alessandro B. Romeo,$^{1}$\thanks{E-mail: romeo@oso.chalmers.se}
        Cathy Horellou$^{1}$ and
        J\"{o}ran Bergh$^{2}$\\
        $^{1}$Department of Astronomy and Astrophysics,
              Centre for Astrophysics and Space Science,
              Chalmers University of Technology,
              \\
              SE-43992 Onsala, Sweden\\
        $^{2}$Department of Mathematics,
              Chalmers University of Technology and G\"{o}teborg University,
              SE-41296 G\"{o}teborg, Sweden}

\begin{document}

\date{Accepted ......... .  Received ......... ; in original form .........}

\pagerange{\pageref{firstpage}--\pageref{lastpage}}

\pubyear{2003}

\maketitle

\label{firstpage}

\begin{abstract}
Noise is a problem of major concern for $N$-body simulations of structure
formation in the early Universe, of galaxies and plasmas.  Here for the first
time we use wavelets to remove noise from $N$-body simulations of disc
galaxies, and show that they become equivalent to simulations with two orders
of magnitude more particles.  We expect a comparable improvement in
performance for cosmological and plasma simulations.  Our wavelet code will
be described in a following paper, and will then be available on request.
\end{abstract}


\begin{figure*}
\scalebox{.80}{\includegraphics{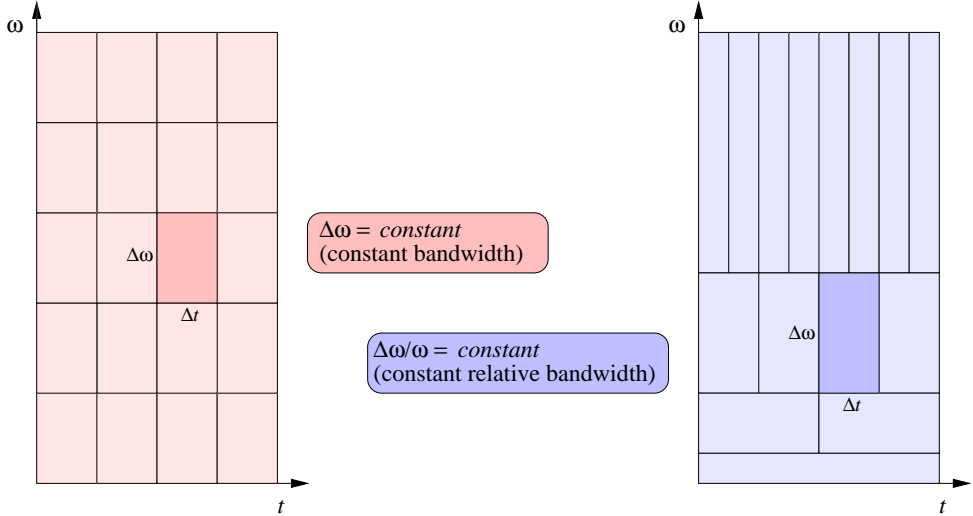}}
\caption{Time-frequency resolution of the windowed Fourier transform (left)
         vs.\ time-frequency resolution of the wavelet transform (right),
         where $\Delta t$ is the time resolution, $\Delta\omega$ is the
         frequency resolution, and $\Delta
         t\cdot\Delta\omega=\mathit{constant}\geq\frac{1}{2}$ (Heisenberg
         uncertainty principle).}
\end{figure*}

\section{INTRODUCTION}

$N$-body simulations of structure formation in the early Universe, of
galaxies and plasmas are limited crucially by noise (e.g., Dawson 1983;
Birdsall \& Langdon 1991; Pfenniger 1993; Pfenniger \& Friedli 1993; Splinter
et al.\ 1998; Baertschiger \& Sylos Labini 2002; Hamana, Yoshida \& Suto
2002; Huber \& Pfenniger 2002; Semelin \& Combes 2002).  The number of
particles $N$ that can be used is several orders of magnitude smaller than
the real number of bodies.  This implies that fluctuations of physical
quantities become drastically exaggerated, and that their contribution to the
overall dynamics can even become dominant.  One way to reduce noise in
simulations is to soften the interaction at short distances.  Softening can
be optimized so as to suppress small-scale fluctuations while respecting the
large-scale dynamical properties (Romeo 1994, 1997, 1998a,\,b; Dehnen 2001;
see also Byrd 1995 and references therein).  On the other hand, noise also
manifests itself as large-scale fluctuations, and the only known way to
reduce them is to increase $N$.  This is very uneconomical because the
computational cost scales at best linearly with $N$.

   Wavelets are a new mathematical tool that has proved to be of great help
for noise reduction in digital signal/image processing (see the beautiful
presentation by Mackenzie et al.\ 2001; see also Bergh, Ekstedt \& Lindberg
1999).  The basic idea behind wavelets is that they provide a
multi-resolution view of the signal.  The signal is analysed first at the
finest resolution consistent with the data, and then at coarser and coarser
resolution levels.  Doing so, wavelets probe the structure of the signal and
the contributions from its various scales.  This is fundamental for noise
reduction, since noise is present on all scales in the signal.  Using
wavelets we can thus remove most of the noise without altering the inherent
structure of the signal.  And we can do this very quickly because the
algorithm is even faster than the fast Fourier transform.

   So why not use wavelets for reducing noise in simulations?  This is indeed
the goal of our paper.  The idea behind such an application is to de-noise
simulations at each timestep.  The resulting improvement is of unprecedented
level compared with classical techniques: wavelet de-noising makes the
simulation equivalent to a simulation with two orders of magnitude more
particles.

   We provide convincing evidence that our wavelet meth\-od works so
successfully by making a detailed prediction and carrying out three hard
tests.  The prediction is based on the improvement in signal-to-noise ratio
for initial models of disc galaxies.  The tests are devoted to probing the
effects of both noise and wavelet de-noising on the simulations, with $N$
spanning two orders of magnitude.  Specifically:
\begin{enumerate}
\renewcommand{\theenumi}{(\alph{enumi})}
   \item The first test concerns the fragmentation of a cool galactic disc,
which is the onset of a gravitational instability (see, e.g., Binney \&
Tremaine 1987).  We show that the time at which the disc fragments is
artificially shortened in simulations with moderate $N$, whereas wavelet
de-noising produces an output comparable to simulations with very large $N$,
as predicted.
   \item The second and third tests concern the heating and accretion
following the fragmentation, respectively.  These are fundamental processes
in the dynamical evolution of disc galaxies, which is induced by
gravitational instabilities via the outward transport of angular momentum and
energy (see, e.g., Binney \& Tremaine 1987).  So these tests also have a
clear physical motivation.  Simulations with moderate $N$ show artificially
enhanced heating and accretion, whereas wavelet de-noising produces an output
consistent with the prediction.
\end{enumerate}
Even though the prediction and the tests are specific to disc galaxies, our
wavelet method is also expected to work successfully when applied to
cosmological and plasma simulations.  This is because the result depends
mainly on the intrinsic properties of wavelets and on the type of noise in
the simulation, as we explain incisively in the discussion.

   The rest of our paper is organized as follows.  The method is presented in
Sect.\ 2, where we give an overview of wavelets and data de-noising using
wavelets, and explain how to use them for de-noising $N$-body simulations.
The prediction is made in Sect.\ 3, the three tests are carried out in Sect.\
4, and further points are discussed in Sect.\ 5.  Finally, the conclusion is
drawn in Sect.\ 6.

\begin{figure*}
\scalebox{.86}{\includegraphics{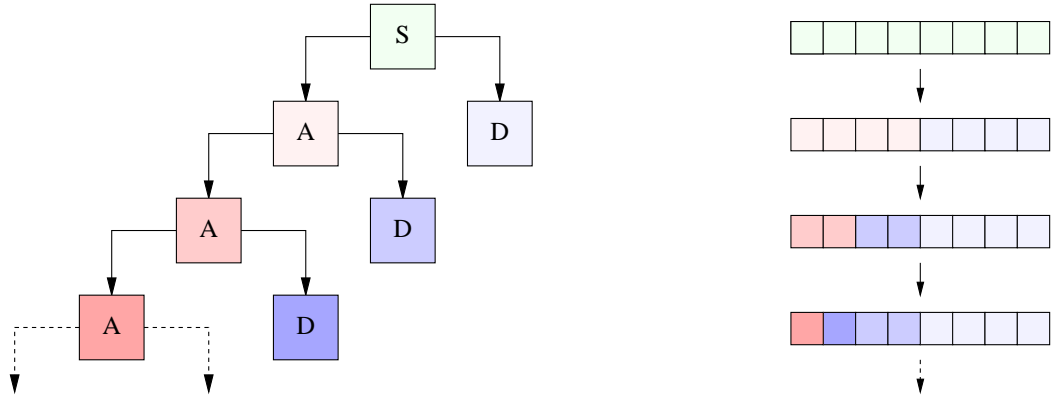}}
\caption{Action of the fast wavelet transform (left) and structure of the
         transformed signal (right), where `S' is the signal, `A' are the
         approximations produced by low-pass filtering and down-sampling, and
         `D' are the details produced by high-pass filtering and
         down-sampling.}
\end{figure*}

\section{METHOD}

\subsection{Wavelets}

Wavelets are a new, very successful, tool for analysing signals, which has
exciting applications in physics, apart from engineering and mathematics
(see, e.g., Mallat 1998; Bergh et al.\ 1999).  In general, real signals are
nonstationary, cover a wide range of frequencies, contain transient
components; and the characteristic frequencies of given segments of the
signal are correlated with their time durations, in the sense that low/high
frequencies imply long/short times.  The standard Fourier analysis is
inappropriate for such signals because it loses all information about the
time localization of a given frequency component%
\footnote{For example, let us consider a musical composition and its Fourier
          transform.  If we interchange various parts of the composition, its
          Fourier transform remains the same but the music becomes different
          or even unrecognizable.},
is very uneconomical and highly unstable with respect to perturbations.  So
real signals require a simultaneous time-frequency analysis%
\footnote{For example, a musical score is a time-frequency analysis of a
          musical composition.}.

   In general, a linear time-frequency transform can be written as
\begin{equation}
s(t)\mapsto S(a,b)=\int_{-\infty}^{\infty}\psi_{ab}^{\ast}(t)s(t)\mathrm{d}t,
\end{equation}
where $s(t)$ is the signal, $S(a,b)$ is the transformed signal at scale $a$
(frequency $\sim 1/a$) and position $b$, and $\psi_{ab}(t)$ is the analysing
function.  In the windowed Fourier transform,
\begin{equation}
\psi_{ab}(t)=\mathrm{e}^{\mathrm{i}t/a}\psi(t-b),
\end{equation}
where the $a$-dependence is a modulation, the $b$-dependence is a
translation, and hence the windows $\psi_{ab}(t)$ have the same width as the
basic $\psi(t)$.  Fig.\ 1 (left) shows that this transform has a fixed
time-frequency resolution.  In the wavelet transform,
\begin{equation}
\psi_{ab}(t)=\frac{1}{\sqrt{a}}\psi\left(\frac{t-b}{a}\right),
\end{equation}
where the $a$-dependence is a dilation ($a>1$) or a contraction ($a<1$), the
$b$-dependence is a translation, and hence the wavelets $\psi_{ab}(t)$ are
self-similar to the basic $\psi(t)$.  Fig.\ 1 (right) shows that such a
transform has an adaptive time-frequency resolution, which is an important
property in favour of its choice.  The inverse wavelet transform is
\begin{equation}
S(a,b)\mapsto s(t)=\frac{1}{c_{\psi}}\int_{-\infty}^{\infty}\mathrm{d}b
\int_{0}^{\infty}\frac{\mathrm{d}a}{a^{2}}\psi_{ab}(t)S(a,b),
\end{equation}
where $c_{\psi}$ is a normalization constant; and, under rather general
assumptions, the admissibility condition for its existence is
\begin{equation}
\int_{-\infty}^{\infty}\psi(t)\mathrm{d}t=0,
\end{equation}
i.e.\ wavelets have zero mean.  A more general requirement is
\begin{equation}
\int_{-\infty}^{\infty}t^{n}\psi(t)\mathrm{d}t=0\;\;\;(n=0,1,\ldots,N),
\end{equation}
i.e.\ wavelets have a certain number $N+1$ of vanishing moments.  This and
other requirements for the choice of $\psi(t)$ are discussed in Sect.\ 2.2.
Summarizing in other words, the continuous wavelet transform has the meaning
of a local filtering, both in time and in scale, and is non-negligible only
when the wavelet matches the signal.

   The continuous wavelet transform can be extended to 2-D signals (images)
by operating not only a scaling and a translation but also a rotation on the
basic wavelet.  And it can also be extended to signals defined on more
general manifolds via group representation theory.

   The wavelet transform can be discretized by setting $a=2^{j}$ and
$b=k\cdot 2^{j}$ ($j,k$ integers), but this condition alone does not
guarantee that the set of wavelets is an orthogonal basis.  Orthogonality is
an important property because it means non-redundancy and thus fast
algorithms, and besides it is a physically useful requirement (see Sect.\
2.2).  Orthogonal and bi-orthogonal wavelet bases can be constructed with a
mathematical technique known as multi-resolution analysis.  The resulting
algorithm is the fast wavelet transform and its complexity is
$4MN_{\mathrm{d}}$, where $M$ is the size of the wavelet and
$N_{\mathrm{d}}=2^{J}$ ($J$ positive integer) is the size of the discrete
signal.  This algorithm is faster than the fast Fourier transform, whose
complexity is $2N_{\mathrm{d}}\log_{2}\!N_{\mathrm{d}}$, and has a data
storage of comparable efficiency.  Fig.\ 2 shows how the fast wavelet
transform acts on the signal.  The signal is decomposed into two parts: an
approximation and a detail.  The approximation is produced by passing the
signal through a low-pass filter and rejecting every other data point.
Analogously, the detail is produced by high-pass filtering and down-sampling.
The approximation is then itself decomposed into two parts: a coarser
approximation and a coarser detail; and the decomposition is iterated.  While
the wavelet analysis consists of decomposing the signal by filtering and
down-sampling, the wavelet synthesis consists of reconstructing the signal by
up-sampling and filtering, and the resulting algorithm is the inverse fast
wavelet transform.

   The fast wavelet transform can be extended to images and $n$-D signals.
The discussion basically follows the 1-D case, except that the signal is
decomposed into $2^{n}$ parts: 1 approximation and $2^{n}-1$ details, one for
each axis and each diagonal; and so on.

\subsection{Data de-noising using wavelets}

Note that in this section `signal', `time' and `frequency' also mean `image',
`space' and `wavevector', or higher-dimensional counterparts, respectively.

\subsubsection{Why do wavelets serve to filter noise so effectively?}

The adaptive time-frequency resolution and the non-redund\-ancy of the fast
wavelet transform have an important practical application: given a noisy
signal, the underlying regular part gets mostly concentrated into few large
wavelet coefficients, whereas noise is mostly mapped into many small wavelet
coefficients.  This means that, if we identify a correct threshold, then we
can set all the small coefficients to zero and get back a signal almost
decontaminated from noise.  This is the idea behind data de-noising and
explains why wavelets serve to filter noise so effectively, independent of
general properties of the data such as the number of dimensions or the
presence of symmetries (see, e.g., Mallat 1998; Bergh et al.\ 1999).

\begin{figure*}
\scalebox{.89}{\includegraphics{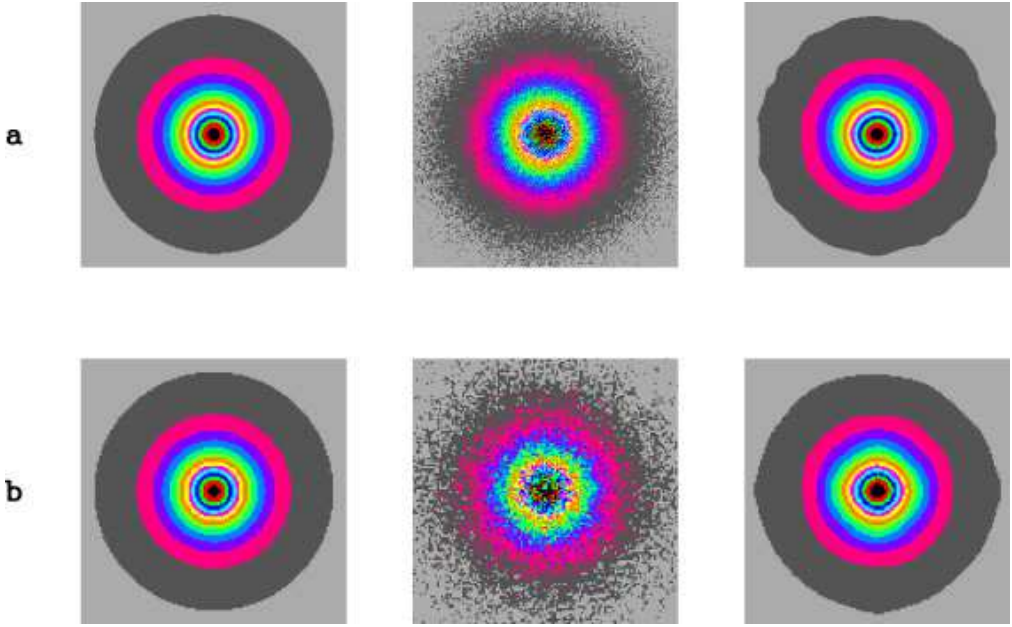}}
\caption{Wavelet de-noising in action.  \textbf{a}: set of models with
         $N=10^{6}$ particles, and with a physical grid of
         $N_{\mathrm{c}}=512\times512$ cells and cell size
         $\Delta_{\mathrm{c}}=0.125$ kpc.  The particle distribution is shown
         for the theoretical model (left), the noisy model (middle) and the
         wavelet de-noised model (right).  In the noisy model the
         signal-to-noise ratio $\mathit{SNR}\simeq9.0$, whereas in the
         wavelet de-noised model $\mathit{SNR}\simeq85.4$.  Thus we predict
         that wavelet de-noised simulations using this initial model are
         equivalent to noisy simulations with approximately 90 times more
         particles, given the Poissonian statistics of noise.  \textbf{b}:
         set of models with $N=10^{5}$, $N_{\mathrm{c}}=256\times256$ and
         $\Delta_{\mathrm{c}}=0.25$ kpc, organized as above (from left to
         right).  In the noisy model $\mathit{SNR}\simeq5.7$, whereas in the
         wavelet de-noised model $\mathit{SNR}\simeq37.2$.  Thus we predict
         that wavelet de-noising improves the performance of the relevant
         simulations by approximately a factor of 42.  Such a prediction is
         tested in Figs 4--6.}
\end{figure*}

\subsubsection{Handling two relevant types of noise}

In the case of Gaussian white noise, a correct threshold can be identified
rigorously if the wavelet basis is orthogonal.  The threshold is proportional
to the standard deviation of noise, which can be robustly estimated from the
smallest-scale detail coefficients, and the proportionality factor depends on
the size of the signal.

   In the case of Poissonian noise, there is a method that produces good
results.  The method is to transform the Poissonian data $Y_{\mathrm{P}}$
into data $Y_{\mathrm{G}}$ with (additive) Gaussian white noise of standard
deviation $\sigma_{\mathrm{G}}=1$:
\begin{equation}
Y_{\mathrm{G}}=2\,\sqrt{Y_{\mathrm{P}}+{\textstyle\frac{3}{8}}}
\end{equation}
(Anscombe 1948), which can then be de-noised as discussed above.
Specifically, the Anscombe transformation has the property to help achieving
additivity, normalization and variance stabilization (Stuart \& Ord 1991).
On the other hand, it has a tendency to fail locally where the data have
small values or large variations (e.g., Kolaczyk 1997; Starck, Murtagh \&
Bijaoui 1998).  Those features may give rise to negative values in the
de-noised data, which can be set to zero.  The Anscombe transformation also
introduces a bias in the data (e.g., Kolaczyk 1997; Starck et al.\ 1998).
The bias is additive and bounded, can be estimated analytically as
\begin{equation}
\mathit{BIAS}\simeq-\frac{1}{4}\left(1-\frac{1}{N_{\mathrm{d}}}\right)
\sigma_{\mathrm{G}}^{2}
\end{equation}
(Stuart \& Ord 1991) or computed numerically and removed from the de-noised
data.

\subsubsection{How should we choose the wavelet?}

In order to optimize data de-noising, we should choose a wavelet that
satisfies the following requirements: (1) it has a small compact support and
is symmetric, for a good time localization; (2) it has a large number of
vanishing moments and is regular, for a good frequency localization; (3) it
is orthogonal, for a correct threshold identification.  On the other hand,
orthogonal wavelets are not symmetric, with one uninteresting exception.  The
best alternative is to choose a symmetric bi-orthogonal wavelet that is also
quasi-orthogonal.  These rules select a few wavelets, from the more than a
hundred commonly used.  The final choice depends on the resolution required
by the data.

\subsection{De-noising of {\boldmath $N$}-body simulations using wavelets}

\subsubsection{How does it work?}

$N$-body simulations commonly use a grid for tabulating the particle density,
and for computing the potential and the field (see, e.g., Hockney \& Eastwood
1988).  The number of particles $n$ in each cell shows fluctuations $|\delta
n|/\langle n\rangle\sim\langle n\rangle^{-1/2}$ with respect to an average
$\langle n\rangle$.  This means that the particle distribution is corrupted
by noise that is basically Poissonian, whereas the noise induced in the
potential and in the field is of a more complex nature.  Using wavelets we
can thus de-noise the particle distribution at each timestep and make the
simulation equivalent to a simulation with many more particles.  This is how
it works.

\subsubsection{Which choices of the wavelet are appropriate?}

A very appropriate choice for de-noising of $N$-body simulations is the
wavelet `rbio\,6.8', described in the documentation of the Matlab Wavelet
Toolbox (Misiti et al.\ 1997).  This wavelet differs significantly from zero
in an interval of approximately three mesh sizes, which is consistent with
the effective spatial resolution of the simulations.  In simulations
dominated by small-scale structures, the wavelet `rbio\,4.4' may be a better
choice.

\begin{figure*}
\scalebox{.88}{\includegraphics{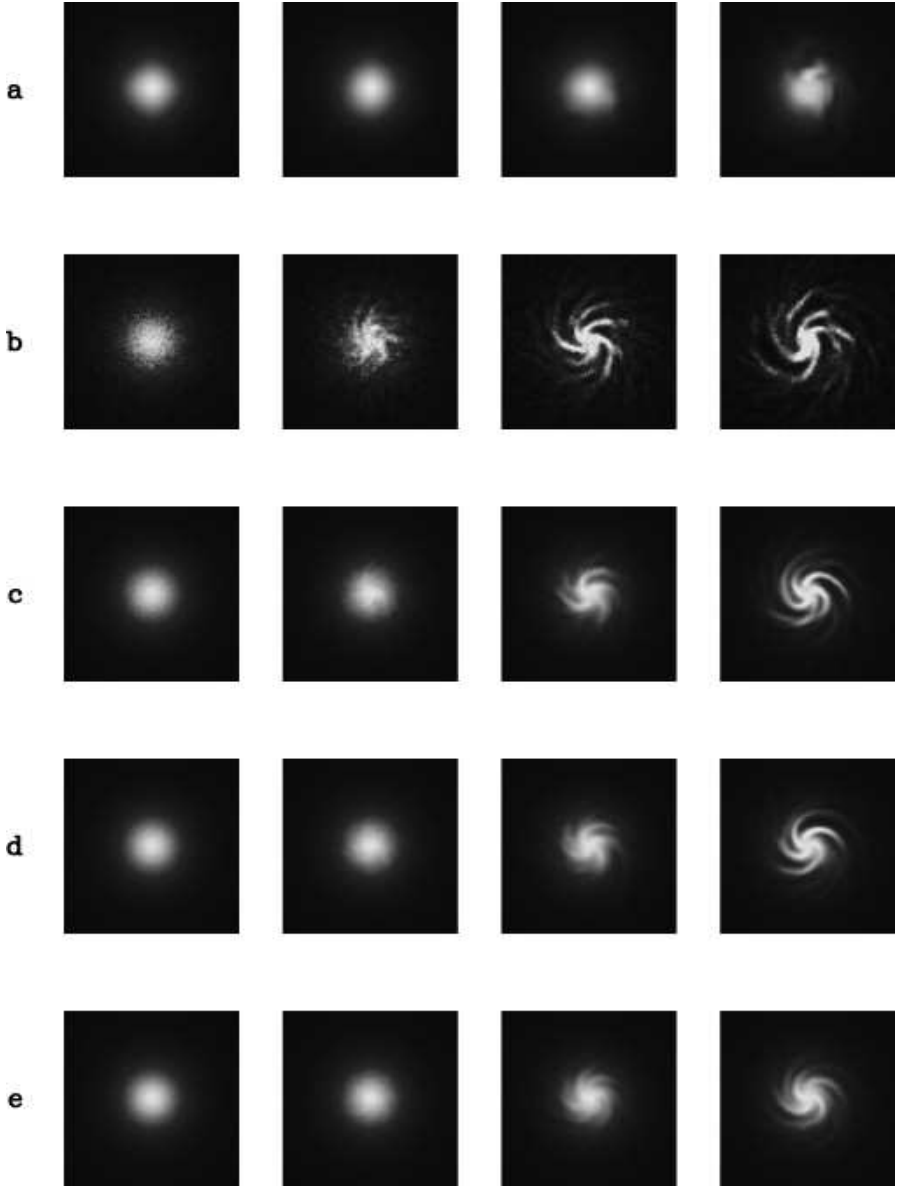}}
\caption{Fragmentation of a cool galactic disc.  \textbf{a}: wavelet
         de-noised simulation with $N=10^{5}$; \textbf{b}: noisy simulation
         with $N=10^{5}$; \textbf{c}: noisy simulation with
         $N=1.8\times10^{6}$; \textbf{d}: noisy simulation with
         $N=4.2\times10^{6}$; \textbf{e}: noisy simulation with
         $N=9\times10^{6}$.  Apart from the number of particles, the initial
         models are the same as in Fig.\ 3b.  For each simulation, the
         particle distribution is shown from 0 Myr to 150 Myr at intervals of
         50 Myr (from left to right).  The time $\tau$ at which the initial
         axial symmetry breaks is a measure of the effect of noise on the
         simulation: a long $\tau$ means a weak effect.  As expected, $\tau$
         increases from \textbf{b} to \textbf{e}.  Note that $\tau$ is longer
         in \textbf{a} than in \textbf{e}.  Thus wavelet de-noising improves
         the performance of the simulation by more than a factor of 90,
         according to the $\tau$-diagnostic.  Such an improvement is well
         beyond the predictions based on the initial models (cf.\ Fig.\ 3b
         and text).}
\end{figure*}

\begin{figure*}
\scalebox{.94}{\includegraphics{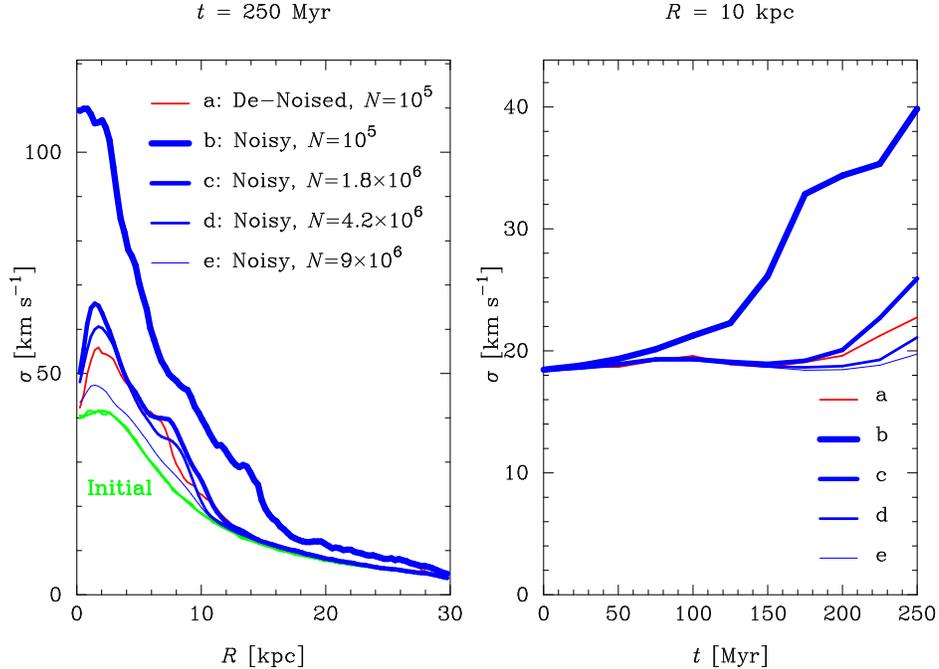}}
\caption{Heating following the fragmentation of a cool galactic disc.  The
         simulations are the same as in Fig.\ 4.  The velocity dispersion
         $\sigma$ is shown as a function of radius $R$ at the initial and
         final times, and as a function of time $t$ at an intermediate
         radius.  The increase of velocity dispersion $\Delta\sigma(R)$ from
         the initial to the final value is a measure of the effect of noise
         on the simulation: a small $\Delta\sigma$ means a weak effect.  In
         all simulations, except b, heating is significant only for $R\la10$
         kpc.  As expected,
         $\Delta\sigma_{\mathrm{b}}>\cdots>\Delta\sigma_{\mathrm{e}}$.  Note
         that $\Delta\sigma_{\mathrm{a}}<\Delta\sigma_{\mathrm{d}}$ for
         $R\la5$ kpc and
         $\Delta\sigma_{\mathrm{a}}\sim\Delta\sigma_{\mathrm{d}}$ with
         oscillations for $R\ga5$ kpc.  Thus wavelet de-noising improves the
         performance of the simulation by a factor of 42 or more, according
         to the $\Delta\sigma$-diagnostic.  This improvement is better than
         predicted with the initial models (cf.\ Fig.\ 3b and text).}
\end{figure*}

\begin{figure}
\centering
\scalebox{.94}{\includegraphics{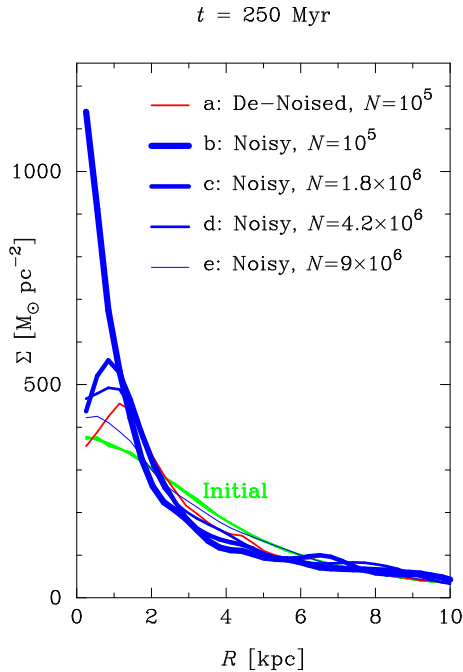}}
\caption{Accretion following the fragmentation of a cool galactic disc.  The
         simulations are the same as in Fig.\ 4.  The mass density $\Sigma$
         is shown as a function of radius $R$ at the initial and final times.
         The peak of final mass density $\hat{\Sigma}$ near the centre gives
         an estimate of the effect of noise on the simulation: a low
         $\hat{\Sigma}$ suggests a weak effect.  As expected,
         $\hat{\Sigma}_{\mathrm{b}}>\cdots>\hat{\Sigma}_{\mathrm{e}}$.  Note
         that $\hat{\Sigma}_{\mathrm{a}}<\hat{\Sigma}_{\mathrm{d}}$.  Thus
         wavelet de-noising improves the performance of the simulation by a
         factor of 42 or more, according to the $\hat{\Sigma}$-diagnostic.
         This improvement is better than predicted with the initial models
         (cf.\ Fig.\ 3b and text).}
\end{figure}

\section{PREDICTION}

In order to make specific predictions, we focus on $N$-body simulations of
disc galaxies, and adopt a widely used Cartesian code (Combes et al.\ 1990).
We choose realistic theoretical models that represent a truncated Kuzmin disc
of scale length $a_{\mathrm{d}}=5$ kpc and cut-off radius
$R_{\mathrm{cut}}=30$ kpc for various $N$, $N_{\mathrm{c}}$ and
$\Delta_{\mathrm{c}}$, where $N$ is the number of particles, $N_{\mathrm{c}}$
is the number of cells in the physical grid and $\Delta_{\mathrm{c}}$ is the
cell size (see, e.g., Binney \& Tremaine 1987).  We then create noisy
simulation models by generating random particle positions and tabulating $n$
at this initial time.  We finally de-noise such models using the wavelet
`rbio\,6.8'.  The amount of noise in the models is quantified by the
signal-to-noise ratio
\begin{equation}
\mathit{SNR}=\left[\frac{{\displaystyle\sum}\,X_{ij}^{2}}{{\displaystyle
\sum}\,(Y_{ij}-X_{ij})^{2}}\right]^{1/2}\,,
\end{equation}
where $X_{ij}$ are the theoretical data and $Y_{ij}$ are either the noisy
data or the wavelet de-noised data.  In the de-noised data, $\mathit{SNR}$
means the inverse of an appropriately defined estimation error.

   Fig.\ 3 illustrates wavelet de-noising in action in a high-resolution case
(Fig.\ 3a) and in a low-resolution case (Fig.\ 3b).  First of all, note that
in the noisy simulation models $\mathit{SNR}\propto\sqrt{\bar{n}}$ for a
given grid size $\sqrt{N_{\mathrm{c}}}\cdot\Delta_{\mathrm{c}}$, where
$\bar{n}=N/N_{\mathrm{c}}$ is the average number of particles per cell.  This
is consistent with the fact that the noise is basically Poissonian.  Then it
follows that wavelet de-noising can improve the signal-to-noise ratio by one
order of magnitude, as if the noisy model had two orders of magnitude more
particles (cf.\ Fig.\ 3a).  Even in noisy models with low signal-to-noise
ratios the improvement is comparable (cf.\ Fig.\ 3b).  The natural prediction
is that the improvement shown for the initial models implies a comparable
improvement in performance for the simulations.

\section{TESTS}

\subsection{Fragmentation of a cool galactic disc}

In order to test such a prediction, we explore a hard problem: the
fragmentation of a cool galactic disc (e.g., Semelin \& Combes 2000; Huber \&
Pfenniger 2001).  A rotating disc with low velocity dispersion is
gravitationally unstable and therefore sensitive to perturbations, which are
amplified and break the initial axial symmetry of the system.  The time that
characterizes symmetry breaking clearly depends on the initial amplitude of
the perturbations, for small perturbations need a long time to grow into an
observable level.  In particular, this is true for the fluctuations imposed
by granular initial conditions.  Thus the symmetry-breaking time is a clear
diagnostic for quantifying the effect of noise on the simulation.  Our
initial models are as in Fig.\ 3b, with the following additional
specifications (irrelevant in that context): disc mass
$M_{\mathrm{d}}=5\times10^{10}$ M$_{\odot}$, Plummer bulge-halo of comparable
mass and scale length, Plummer softening with softening length $s=0.25$ kpc,
Safronov-Toomre parameter $Q=0.7$ (see, e.g., Binney \& Tremaine 1987).  We
vary $N$ so as to test in detail the improvements predicted by Fig.\ 3b, and
by an analogous examination of the gravitational field of the disc (a factor
of 18).  We run the simulations for 250 Myr, a typical dynamical time.

   Fig.\ 4 illustrates that in the wavelet de-noised simulation the
symmetry-breaking time is even longer than in the best noisy simulation,
which uses the largest number of particles allowed by computer memory.  This
is a strong piece of evidence that wavelet de-noising outperforms the
predictions.

\subsection{Heating following the fragmentation of a cool galactic disc}

As an additional test, we investigate the heating following the fragmentation
of a cool galactic disc.  When spiral gravitational instabilities reach a
sufficiently large amplitude, the velocity dispersion of the disc starts to
increase by collective relaxation (e.g., Zhang 1998; Griv, Gedalin \& Yuan
2002).  The heat produced in a dynamical time is low if the initial amplitude
of the instabilities is small.  Therefore the increase of velocity dispersion
is another diagnostic for quantifying the effect of noise on the simulation.

   Fig.\ 5 illustrates that, on the whole, in the wavelet de-noised
simulation the increase of velocity dispersion is smaller than in the noisy
simulation with 42 times more particles.  This is a further piece of evidence
that wavelet de-noising outperforms the predictions.

\subsection{Accretion following the fragmentation of a cool galactic disc}

As a further test, we analyse the accretion following the fragmentation of a
cool galactic disc.  The amplification of spiral gravitational instabilities
produces not only heating but also re-distribution of matter in the disc,
which appears more evidently as accretion near the centre (e.g., Zhang 1998;
Griv et al.\ 2002).  The mass accreted in a dynamical time is low if the
initial amplitude of the instabilities is small.  So the peak of mass density
is still another diagnostic for quantifying the effect of noise on the
simulation.

   Fig.\ 6 strengthens our result by confirming, once again, that wavelet
de-noising outperforms the predictions.

\section{DISCUSSION}

Wavelet de-noising suppresses almost all the noise but introduces slight
local biases, which have no significant effect on the conservation of angular
momentum and energy.  In fact, the deviations are less than 0.04\% and 0.06\%
per dynamical time, respectively, and compare well with those typical of the
code (Combes et al.\ 1990).  The improvement in performance concerns not only
the fidelity of simulations to real systems, shown in Figs 3--6 and discussed
above, but also the standard computational issues (time, memory and storage).
This improvement is by one/two orders of magnitude in the low/high-resolution
case examined.  Note in this context that an important, albeit non-classical,
reason for increasing the number of particles is to improve the resolution.
Optimal choices of $N$, $\Delta_{\mathrm{c}}$ and $s$ are interdependent
(e.g., Hernquist, Hut \& Makino 1993; Dehnen 2001), and besides they strongly
depend on the physical problem under investigation (Romeo 1998a; Dehnen
2001).  Wavelet de-noising acts so as to increase the equivalent number of
particles, therefore analogous considerations apply.  Further points are
discussed below.

\subsection{Large vs.\ small spatial scales}

The spatial scales on which noise is `significant' depend upon what physical
quantity is considered: the particle density, the potential, the field, or
other quantities related to the dynamical effect of noise such as the
relaxation time (in this case one often refers to the relative contribution
of distant and close encounters; e.g., Dehnen 2001).  In particular, the
particle distribution is affected by noise that is basically Poissonian and
white, and hence significant on all scales.  As noted in Sect.\ 2.2 and
adapted to the present context, wavelets succeed in suppressing this type of
noise not only on small but also on \emph{large} scales, \emph{without}
damping real instabilities.  Indeed, this is what makes wavelets so
successful in comparison with other methods of noise reduction (see the
literature quoted).

\subsection{3D vs.\ 2D, and initial (a)symmetry}

The prediction and the tests shown in previous sections concern disc
galaxies, whose simulation models are commonly flat and initially
axisymmetric.  Appropriate 3-D tests regarding this or other types of
galaxies would demand an exceedingly large number of particles,
$N\sim10^{8}\mbox{--}10^{9}$, in order to have a satisfactory signal-to-noise
ratio in the basic simulation (an average of a few particles per cell) and in
order to be able to run simulations with two orders of magnitude more
particles.  On the other hand, as noted in Sect.\ 2.2.1, the remarkable
efficiency of wavelets in suppressing noise is an \emph{intrinsic} property
of such functions, which has been tested \emph{extensively} in the
literature.  Because of that the initial symmetry is also irrelevant, apart
from the fact that it breaks quickly.  If any remark should be made, then
note that axisymmetry is an unfavourable initial condition since the fast
wavelet transform has privileged directions: the Cartesian axes and diagonals
(see Sect.\ 2.1).  Thus the result of our paper is also expected to hold in
3D, with or without initial symmetry.

\subsection{Plasma and cosmological vs.\ galaxy simulations}

Can our wavelet method be applied to plasma and cosmological simulations?

   The case of plasma simulations is rather clear.  Even if the initial
models are non-noisy, Poissonian noise develops in few dynamical times (see,
e.g., Dawson 1983; Birdsall \& Langdon 1991).  Wavelet de-noising can be
applied from the start, since it prevents the onset of noise without
affecting the quiet model.

   The case of cosmological simulations is more complex but is also quite
clear.  The initial conditions consist of setting up a non-noisy uniform
particle distribution, and of imposing small random fluctuations with
Gaussian statistics and a given power spectrum (e.g., Efstathiou et al.\
1985; Sylos Labini et al.\ 2002).  Poissonian noise develops naturally, as a
reaction of the system to the initial order and hence reduced entropy.  The
onset of Poissonian noise is especially quick in cold-dark-matter
simulations, where structures form bottom-up and the first virialized systems
contain a small number of particles (e.g., Binney \& Knebe 2002).  There is a
natural way to apply our wavelet method to cold-dark-matter simulations.  It
is to de-noise them only over a range of scales that is adapted to the phase
of clustering: from the cell size to the size of the structures that have
formed latest.  Analogous ideas can be implemented in the context of other
cosmological models.  In this way wavelet de-noising prevents the onset of
Poissonian noise without affecting the imposed fluctuations, or the quiet
background.

   Thus our wavelet method can be applied not only to galaxy but also to
cosmological and plasma simulations, with minor adaptions.

\section{CONCLUSION}

The conclusion of this paper is that wavelet de-noising allows us to improve
the performance of $N$-body simulations up to two orders of magnitude.  This
result has been shown for simulations of disc galaxies, and can be
generalized to other grid geometries and particle species than those used
here.  Besides, wavelet de-noising can be adapted for a variety of
constraints (e.g., partial de-noising at given scales or at a pre-assigned
level) and initial conditions (e.g., partially noisy or quiet starts), and
thus have important applications also in cosmological and plasma simulations.

   Our wavelet code will be described in a following paper, and will then be
available on request.

\section*{ACKNOWLEDGMENTS}

It is a great pleasure to thank Gene Byrd, Fran\c{c}oise Combes, Stefan
Goedecker and Daniel Pfenniger for strong encouragement, valuable suggestions
and discussions.  We are very grateful to John Black, Alan Pedlar and Gustaf
Rydbeck for strong support.  We also acknowledge the financial support of the
Swedish Research Council and a grant by the `Solveig och Karl G Eliassons
Minnesfond'.

\bsp

\label{lastpage}

\end{document}